\documentclass[twocolumn,showpacs,preprintnumbers,superscriptaddress,amsmath,amssymb]{revtex4}

\usepackage{graphicx}
\usepackage{dcolumn}
\usepackage{bm}

\begin{document}

\preprint{APS/123-QED}

\title{Evidence for suppressed metallicity on the surface of La$_{2-x}$Sr$_x$CuO$_4$ and Nd$_{2-x}$Ce$_x$CuO$_4$}

\author{M. Taguchi}

\affiliation{Soft X-ray Spectroscopy Lab, RIKEN/SPring-8, Mikazuki, Sayo, Hyogo 679-5148, Japan}

\author{A. Chainani}

\affiliation{Soft X-ray Spectroscopy Lab, RIKEN/SPring-8, Mikazuki, Sayo, Hyogo 679-5148, Japan}

\author{K. Horiba}

\affiliation{Soft X-ray Spectroscopy Lab, RIKEN/SPring-8, Mikazuki, Sayo, Hyogo 679-5148, Japan}

\author{Y. Takata}

\affiliation{Soft X-ray Spectroscopy Lab, RIKEN/SPring-8, Mikazuki, Sayo, Hyogo 679-5148, Japan}

\author{M. Yabashi}

\affiliation{Coherent X-ray Optics Lab, RIKEN/SPring-8, Mikazuki, Sayo,
 Hyogo 679-5148, Japan}
\affiliation{JASRI/SPring-8, Mikazuki, Sayo, Hyogo 679-5148, Japan}

\author{K. Tamasaku}

\affiliation{Coherent X-ray Optics Lab, RIKEN/SPring-8, Mikazuki, Sayo, 
Hyogo 679-5148, Japan}

\author{Y. Nishino}

\affiliation{Coherent X-ray Optics Lab, RIKEN/SPring-8, Mikazuki, Sayo, 
Hyogo 679-5148, Japan}

\author{D. Miwa}

\affiliation{Coherent X-ray Optics Lab, RIKEN/SPring-8, Mikazuki, Sayo, 
Hyogo 679-5148, Japan}

\author{T. Ishikawa}

\affiliation{Coherent X-ray Optics Lab, RIKEN/SPring-8, Mikazuki, Sayo, 
Hyogo 679-5148, Japan}

\author{T. Takeuchi}

\affiliation{Soft X-ray Spectroscopy Lab, RIKEN/SPring-8, Mikazuki, Sayo, Hyogo 679-5148, Japan}

\author{K. Yamamoto}

\affiliation{Soft X-ray Spectroscopy Lab, RIKEN/SPring-8, Mikazuki, Sayo, Hyogo 679-5148, Japan}

\author{M. Matsunami}

\affiliation{Soft X-ray Spectroscopy Lab, RIKEN/SPring-8, Mikazuki, Sayo, Hyogo 679-5148, Japan}

\author{S. Shin}

\affiliation{Soft X-ray Spectroscopy Lab, RIKEN/SPring-8, Mikazuki, Sayo, Hyogo 679-5148, Japan}
\affiliation{Institute for Solid State Physics, University of Tokyo, Kashiwa, Chiba 277-8581, Japan}

\author{T. Yokoya}

\affiliation{JASRI/SPring-8, Mikazuki, Sayo, Hyogo 679-5148, Japan}

\author{E. Ikenaga}

\affiliation{JASRI/SPring-8, Mikazuki, Sayo, Hyogo 679-5148, Japan}

\author{K. Kobayashi}

\affiliation{JASRI/SPring-8, Mikazuki, Sayo, Hyogo 679-5148, Japan}

\author{T. Mochiku}

\affiliation{Superconducting Materials Center, National Institute for Materials Science, Tsukuba, Ibaraki 305-0047, Japan}

\author{K. Hirata}

\affiliation{Superconducting Materials Center, National Institute for Materials Science, Tsukuba, Ibaraki 305-0047, Japan}

\author{J. Hori}

\affiliation{Department of Quantum Matter, ADSM, Hiroshima University, Higashi-Hiroshima, Hiroshima 739-8530, Japan}

\author{K. Ishii}

\affiliation{Department of Quantum Matter, ADSM, Hiroshima University, Higashi-Hiroshima, Hiroshima 739-8530, Japan}

\author{F. Nakamura}

\affiliation{Department of Quantum Matter, ADSM, Hiroshima University, Higashi-Hiroshima, Hiroshima 739-8530, Japan}

\author{T. Suzuki}

\affiliation{Department of Quantum Matter, ADSM, Hiroshima University, Higashi-Hiroshima, Hiroshima 739-8530, Japan}

\date{\today} 

\begin{abstract}
Hard X-ray Photoemission spectroscopy (PES) of copper core electronic states, with a probing depth of $\sim$60 \AA, is used to show that the Zhang-Rice singlet feature is present in La$_2$CuO$_4$ but is absent in Nd$_2$CuO$_4$. Hole- and electron doping in La$_{2-x}$Sr$_x$CuO$_4$ (LSCO) and Nd$_{2-x}$Ce$_x$CuO$_4$ (NCCO) result in new well-screened features which are missing in soft X-ray PES. Impurity Anderson model calculations establish metallic screening as its origin, which is strongly suppressed within 15 $\text{\AA}$  of the surface. Complemented with X-ray absorption spectroscopy, the small chemical-potential shift in core levels ($\sim0.2$ eV) are shown to be consistent with modifications of valence and conduction band states spanning the band gap ($\sim1$ eV) upon hole- and electron-doping in LSCO and NCCO. 
\end{abstract}

\pacs{74.72.Dn, 74.72.Jt, 78.20.Bh, 79.60.-i}

\maketitle

Hole- and electron-doping by chemical substitutions in single layer copper-oxides (as in La$_{2-x}$A$_x$CuO$_4$, A=Ba,Sr and Nd$_{2-x}$Ce$_x$CuO$_4$) transforms an antiferromagnetic insulator to an exotic metal with superconductivity\cite{ima98}. The properties of hole- and electron- doped high-Tc cuprate are determined by electronic states near the chemical-potential\cite{ima98}, accompanied with characteristic features in core levels\cite{cox92}. Soft X-ray (SX, $h\nu$$\sim$1000-1500 eV) core level photoemission spectroscopy (PES) with a probing depth of $\sim$10-15 $\text{\AA}$ is valuable in studying valence change, chemical-potential-shift and screening effects in solids\cite{bri90}. Combination of core level PES with model calculations have been used to describe the parent insulating cuprates La$_2$CuO$_4$ (LCO) and Nd$_2$CuO$_4$ (NCO) as charge-transfer insulators in the Zaanen-Sawatzky-Allen classification scheme\cite{zaa85}, with the on-site Coulomb energy ($\approx$8 eV), being much larger than the charge transfer energy ($\approx$2 eV) between the O $2p$ and Cu $3d$ states\cite{fuj87,she87,cum93,koi02}. 

La$_{1.85}$Sr$_{0.15}$CuO$_4$ (LSCO) and Nd$_{1.85}$Ce$_{0.15}$CuO$_4$ (NCCO) are prototypical of hole- and electron-doped cuprates and exhibit a $d_{x^2-y^2}$ superconducting gap. The normal phase resistivity ($\rho$$\propto$$T^2$) is like a Fermi-liquid for NCCO\cite{tsu89} but non-Fermi-liquid-like ($\rho$$\propto$$T$) for LSCO\cite{gur87}. The strong correlations lead to special spectral behaviour such as non-local screening effects\cite{vee94}, and anomalous spectral weight transfer upon doping\cite{esk91}. While valency and chemical potential changes in the high-Tc cuprates can be probed with SX-PES, in spite of several core level and valence band PES studies, there remains a seemingly simple and yet unresolved puzzle about the doping dependent electronic structure of the superconducting cuprates\cite{fuj87,she87,cum93,koi02,vee94,esk91,ste03,all90,har01}. The puzzle involves distinguishing between 'mid-gap pinning' or 'crossing the gap' scenario to simultaneously explain changes in core levels and valence bands. The mid-gap pinning scenario\cite{fuj87,she87,all90,har01} involves formation of new states within the band gap on hole- and electron-doping. It explains the small chemical potential shift of -0.2 eV (or +0.2 eV) in O $1s$ core levels PES of LSCO (or NCCO) compared to undoped LCO (or NCO), but is inconsistent with the large optical gap onset ($\sim$1.0 eV) of the insulating parents\cite{uch91}. In an alternative picture, the chemical potential moves to the top of the valence band by hole-doping and bottom of the conduction band on electron-doping. Using resonant PES\cite{ste03}, it was shown that electron- and hole-doping leads to a crossing of the gap ($\sim$1.0 eV) from NCCO to LSCO. However, the small chemical potential shift in O $1s$ core levels cannot be explained by this scenario.

\begin{figure}
\includegraphics[scale=.55]{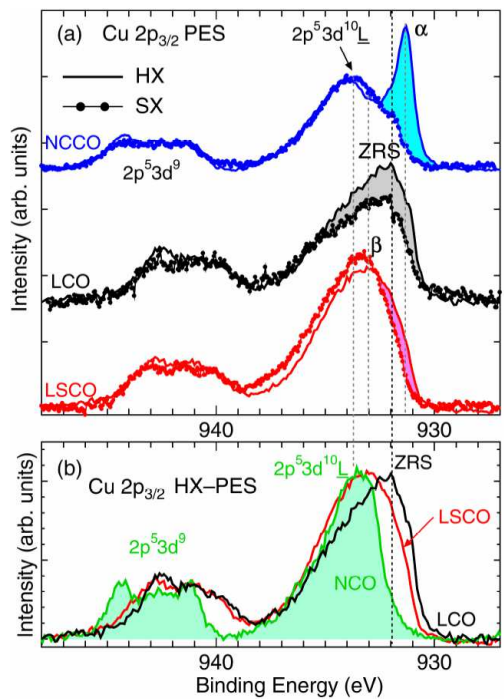}
\caption{\label{fig1}   
Experimental Cu $2p$ HX-PES spectra. (a), Comparison between HX-PES (solid line) and SX-PES (line with symbols) for electron-doped NCCO, undoped LCO and hole-doped LSCO. Shaded regions highlight the differences in HX-PES. (b), HX-PES comparison of undoped NCO, LCO and hole-doped LSCO. }
\vspace{-0.25in}
\end{figure}

While many SX-PES of the Cu $2p$ core levels of LSCO and NCCO have been performed, the spectra show very little change in binding energy and negligible difference in spectral shape upon doping\cite{fuj87,she87,cum93,koi02,har01}. This leads to another significant issue: the presence of the predicted Zhang-Rice singlet (ZRS) in Cu $2p$ core levels of the insulating cuprates\cite{vee94}, which is considered very important for superconductivity but has not been observed to date by core level PES. However, recent spin-polarized resonant valence band PES studies indicate ZRS states closest to the chemical potential for undoped cuprates\cite{ghi02,tje03}. Significantly, these studies distinguish between LSCO and NCCO: the ZRS survives in LSCO\cite{ghi02} but is absent in NCCO\cite{tje03}. These puzzles bring into question the role of depth-sensitivity of PES, which has often led to controversies regarding surface versus bulk electronic structure. Although pioneering core level PES with a photon energy of 8 keV (probing depth $\sim$80 $\text{\AA}$) has been reported 30 years ago\cite{lin74}, its importance for separating the surface and bulk electronic structure has been recognized only recently\cite{dal01}. 

\begin{figure}
\includegraphics[scale=.41]{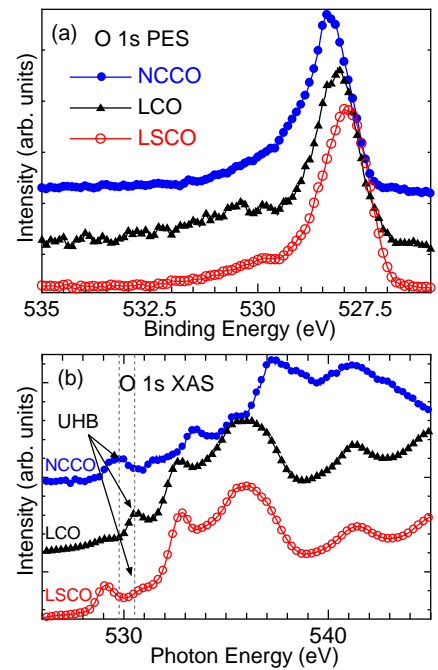}
\caption{\label{fig2}     
(Color online) O $1s$ HX-PES spectra (a) and O $1s$ XAS spectra (b) of NCCO, LCO and LSCO. The relative position of the UHB with respect to the O $1s$ core level in LSCO and NCCO can be estimated from the O $1s$ XAS. The O $1s$ XAS shows a "pre-peak" feature due to the UHB that is brought about by O $2p$-Cu $3d$ hybridization.}
\vspace{-0.25in}
\end{figure}

In an attempt to describe all the spectroscopic physics correctly, we carry out core level Hard X-ray (HX) PES\cite{tam01,kob03} of single-crystal electron-doped NCCO, hole-doped LSCO, and undoped NCO and LCO at BL29 of SPring-8\cite{tam01}. The kinetic energy of the Cu $2p$ core level corresponds to a probing depth of $\sim$60 $\text{\AA}$ as determined by the inelastic mean free path\cite{bri90}. 
Single crystals of NCO, NCCO, LCO and LSCO were grown by the travelling solvent floating zone method. NCCO and LSCO showed a superconducting T$_c$ of 22 K and 36 K, respectively. HX-PES was performed using a photon energy $h\nu$=5.95 keV, at a vacuum of $1\times 10^{-10}$ Torr. The measurements were carried out at undulator beam line BL29XU, SPring-8 using a Scienta R4000-10KV electron analyzer. The energy width of incident X-rays was 70 meV, and the total energy resolution,  $\Delta$E was set to $\sim$0.4 eV. SX-PES ($h\nu$=1500 eV) was performed at BL17SU, with  $\Delta$E$\sim$0.3 eV. All measurements used a normal emission geometry to maximize depth sensitivity. Sample temperature was controlled to $\pm $ 2 K during measurements. All samples were fractured in-situ and NCCO and LSCO were measured at 35 K while LCO and NCO were measured at room temperature. Gold $4f$ core levels were measured to calibrate the energy scale.

Figure~1(a) shows Cu $2p_{3/2}$ HX-PES spectra and SX-PES ($h\nu$=1.5 keV) of NCCO, LCO and LSCO. Figure~1(b) shows the Cu $2p_{3/2}$ HX-PES spectra of undoped NCO, LCO and hole-doped LSCO. The NCO Cu $2p_{3/2}$ spectrum consists of a main peak at 933.5 eV ($2p^53d^{10}\underline{L}$ state: $\underline{L}$ represents the ligand hole) and a broad satellite centred at 943 eV ($2p^53d^9$ state), and is very similar to earlier SX-PES\cite{cum93,koi02}. The HX-PES spectra of NCCO, LCO and LSCO (Figs.~1(a) and 1(b)) are clearly different and provide new results:

(i)	The LCO HX-PES spectrum (Fig.~1(b)) shows a main peak at $\sim$932 eV and an additional shoulder at $\sim$933.5 eV binding energy. Using a multi-site cluster model\cite{vee94,oka97}, it was shown that even for the insulating parent, the Cu $2p_{3/2}$ spectrum has a low binding energy signature of the ZRS due to non-local screening, while the $2p^53d^{10}\underline{L}$ state occurs at higher binding energy. But earlier SX-PES showed only a single peak at 933.5 eV due to the $2p^53d^{10}\underline{L}$ state. HX-PES clearly shows that, in LCO, the peak at 933.5 eV is the $2p^53d^{10}\underline{L}$ and the new feature at $\sim$932 eV is the ZRS peak. The present result is the first observation of the ZRS feature in Cu $2p$ core level PES of LCO. The ZRS feature is missing in the NCO HX-PES spectrum (Fig.~1(b)).

(ii) The LSCO spectrum (Fig.~1(b)) shows clear changes compared to LCO. The ZRS feature is retained on hole-doping, but is weakened compared to LCO, and additional spectral weight is seen at higher binding energy (feature $\beta$). Since the main peak width (nearly 4 eV FWHM) is very large in LSCO, it consists of more than a single configuration ($2p^53d^{10}\underline{L}$ feature and feature $\beta$). While the doping is supposed to cause a valency change (Cu$^{3+}$ content due to hole-doping), it is impossible to separate out Cu$^{3+}$ and retained ZRS features, from the $2p^53d^{10}\underline{L}$ feature. Within experimental accuracy, the $2p^53d^{10}\underline{L}$ feature in all the materials occurs at the same position ($\pm $0.1 eV) as in SX-PES data\cite{fuj87,she87,cum93,koi02}. 

\begin{table}
\begin{tabular}{ccccccccc}
\multicolumn{8}{c}{}\\
\hline
& $\Delta$ & $\Delta^*$ & $U_{dd}$  & $U_{dc}(2p)$ & $V(e_g)$ & $V^*(e_g)$ & $T_{pp}$ & \\
\hline
 $\text{NCCO}$ & 3.0 & 0.25 & 8.0 & 10.5 & 3.5 & 1.8 & 1.0 & \\
 $\text{LSCO}$ & 3.6 & 1.35 & 8.0 & 10.0 & 3.75 & 1.25 & 1.0 & \\
\hline

\end{tabular}
\caption{Estimated parameter values for NCCO and LSCO}
\vspace{-0.25in}
\end{table}

(iii) The HX-PES spectrum for NCCO (Fig.~1(a)) shows a sharp low binding energy feature $\alpha$ compared to NCCO SX-PES and undoped NCO HX-PES spectra. This feature $\alpha$ was not observed in earlier (and also present) SX-PES studies\cite{ste03}. Its energy position is different from the ZRS feature in LCO. More importantly, since the ZRS feature is missing in undoped NCO, its origin is different and discussed in the framework of Impurity Anderson model (IAM) calculations later. For the same sample and surface preparation (single crystal cleaved surfaces), SX-PES shows a peak at $\sim$933.5 eV and a weak shoulder at $\sim$932.0 eV. The SX-PES is very similar to a recent report, with the $2p^53d^{10}\underline{L}$ feature at $\sim$933.5 eV and the weak shoulder was attributed to the $2p^53d^{10}$ state\cite{ste03}. In terms of the Ce content (x=0.15), a maximum of 15 \% of the spectral intensity can arise due to the formally Cu$^{1+}$ $3d^{10}$ configuration, in contrast to the observed intensity ($\sim$30 \%). This rules out a simple $3d^{10}$ configuration interpretation. 

(iv) The 'ZRS', '$\alpha$' and '$\beta$' features in LCO, NCCO and LSCO (Fig.1~(a)) are clearly observed in HX-PES. While these features are missing in earlier SX-PES\cite{fuj87,she87,cum93,koi02,har01}, we find evidence for the ZRS feature in LCO and broadening in LSCO even from our SX-PES measurements carried out on high-quality single crystals fractured in-situ. Also, the satellite features between NCCO and NCO show little change with doping, and so also for LSCO and LCO. However, a large shift of nearly 1.5 eV to higher binding energy is observed for the NCO/NCCO satellite compared to LCO/LSCO.

While the Cu $2p$ HX-PES spectra are significantly different compared to SX-PES spectra, the O $1s$ core level HX-PES spectra are very similar to the SX-PES spectra (Fig.~2(a)). The O $1s$ levels show a shift in peak position towards higher binding energy (+0.25 eV) in the electron-doped NCCO and towards lower binding energy in the hole-doped LSCO (-0.2 eV), in accord with earlier SX-PES studies\cite{cum93,koi02,har01}. Generally, it is believed that these peak shifts may reflect the chemical potential shift. 

\begin{figure}
\includegraphics[scale=.48]{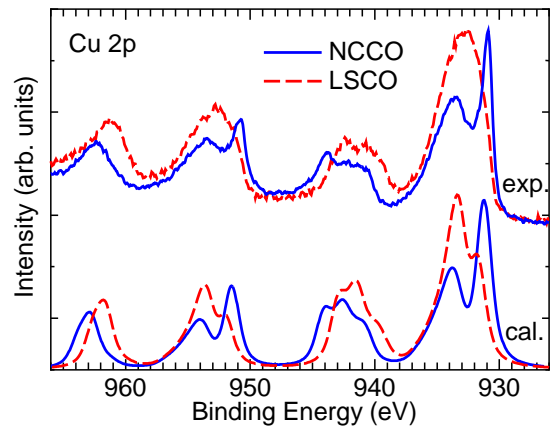}
\caption{\label{fig3}   
(Color online) IAM calculations for the Cu $2p$ core level PES of NCCO and LSCO (lower panel) compared with experimental spectra (upper panel). The broad satellite around 944 eV is due to multiplets of $2p^53d^9$ character. The O $2p$ band width broadens the $2p^53d^{10}\underline{L}$ peak around 933.5 eV.}
\vspace{-0.25in}
\end{figure}

To understand the origin of the doping dependent features in Cu $2p$ spectra, we have performed IAM calculations in the D$_{4h}$ local symmetry including intra-atomic multiplets. Here we retain only a single Cu atom (core-hole site) and allow charge transfer between the Cu $3d$ state and the O $2p$ band as well as the Cu $3d$ state and doping-induced states. The essential new feature is the charge transfer from doping-induced states to the upper Hubbard band (UHB) $\Delta^*$ defined as $E(3d^{10}\underline{C})-E(3d^9)$. 
The usual charge transfer energy (from O $2p$ band to UHB), is defined as $E(3d^{10}\underline{L}) - E(3d^9)$. The $3d^{10}\underline{C}$ represent the charge transfer between Cu $3d$ state and the doping-induced state at E$_F$. The O $2p$ bands and doping-induced states $\varepsilon_k$ are approximated by $N$ discrete levels and the $k$ dependence of the hybridization is assumed to be elliptical. (The technical details of the calculation have been reported in Ref.\cite{sei90,hor04,tag04}.)

The calculated results are shown in Fig.~3 with experimental results of LSCO and NCCO. The calculations reproduce well the main peaks and satellite structure. The sharp peak at low binding energy in NCCO originates from core-hole screening by doping-induced states at E$_{F}$, the $2p^53d^{10}\underline{C}$ state. The obtained parameter values show small differences for LSCO and NCCO as summarized in Table I. The most important parameter is  $\Delta^*$, which represents the energy difference between the UHB and doping-induced states. The small value of $\Delta^*$(=0.25 eV) for NCCO indicates that the doping-induced states lie just below the UHB, whereas a large $\Delta^*$(=1.35 eV) of LSCO describes the situation for doping-induced states lying near the top of the valence band, with the UHB separated by $\Delta^*$. However, this still does not explain the small chemical potential shift of O $1s$ levels (+0.25 to -0.2 eV) from NCCO to LSCO. Using O $1s$ XAS to probe the unoccupied density of states, it is known that hole-doping in LSCO develops a new feature\cite{che91} below the UHB while electron-doping in NCCO results\cite{rom91} in effectively enhancing the intensity of the UHB itself (Fig.~2(b)). The results indicate that the onset peak position of the UHB for LSCO is 0.8-1.0 eV higher in energy than that of NCCO with respect to O $1s$ core level\cite{che91,rom91}.

Putting the O $1s$ HX-PES, O $1s$ XAS and Cu $2p$ HX-PES data together in an energy level diagram, we arrive at the picture (Fig.~4) describing all the spectroscopic results: (i) The UHB of LSCO is higher in energy than that of NCCO, following the O $1s$ XAS. (ii) The difference in  $\Delta^*$ is an approximate measure of the band gap of both NCO and LCO ($\sim$1.0 eV) which is consistent with the resonant PES\cite{ste03} and optical gap\cite{uch91}. (iii) The chemical potential shift between NCCO and LSCO is rather small compared with the optical gap, explaining the small peak shift in O $1s$ spectra compared to a 'crossing the gap' picture. (iv) Figure~4 clearly shows that electron-doping induced states lie at or near the bottom of the UHB in NCCO while the hole-doping induced states are situated near the top of the valence band, separated by $\sim$1 eV. It is important to note that, in the mid-gap pinning scenario, $\Delta^*$ has to be the same value in both electron- and hole-doped systems and would result in the same spectral shape of Cu $2p$ spectra for NCCO and LSCO. 

\begin{figure}
\includegraphics[scale=.40]{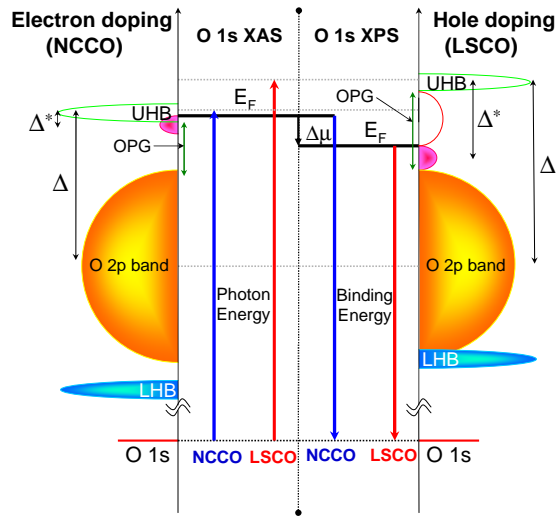}
\caption{\label{fig4}   
(Color online) Schematic illustration of the energy levels of LSCO and NCCO obtained from the IAM analysis. OPG represents the optical gap in undoped materials. The Fermi level (E$_F$) separates the occupied density of states (shaded regions) from the unoccupied density of states.}
\vspace{-0.25in}
\end{figure}

The importance of well-screened and unscreened features in core level spectra to study the Mott metal-insulator transition in correlated oxides was first proposed by Cox $et$ $al.$ for a series of ruthenium oxides\cite{cox92,cox83}. The present study with a probing depth of 60 $\text{\AA}$ using $\sim$6 keV photons distinguishes between metallic and ligand screening, and shows metallic screening is strongly suppressed in the top 15 $\text{\AA}$ of the copper oxides. Core-level HX-PES thus provides unequivocal information not readily obtained by SX-PES. A recent dynamical mean-field theory (DMFT)\cite{kim04} confirms the conclusions of Cox $et$ $al.$ The present theoretical model may be viewed as a simplified IAM which qualitatively reproduces the DMFT results for a single band Hubbard model\cite{kim04} as well as multi-site cluster calculation\cite{vee94,oka97,tag04}. In addition, it includes atomic multiplets which are important for the structure in the Cu $2p$ satellites here, as well as the main peak and satellite seen in other correlated oxides which show doping and temperature dependent well-screened features\cite{hor04,tag04}.

In conclusion, bulk sensitive HX-PES is used to show that LSCO and NCCO exhibit new bulk character electronic states originating in metallic screening, which are strongly suppressed within $\sim$15 $\text{\AA}$ of the surface. IAM calculations, complemented with O $1s$ XAS, explain the intriguingly small chemical-potential-shift in core levels ($\sim$0.2 eV for hole- or electron-doping in LSCO and NCCO) as well as valence and conduction band modifications spanning the band gap ($\sim$1 eV). Electronic structure studies of the subsurface region ($\sim$20-100 $\text{\AA}$) of solids, with important applications in corrosion science, ultrashallow semiconductor devices, interfaces, buried layers, etc. become possible with HX-PES.

This work was partially supported by the Ministry of Education, Science, Sports and Culture through a Grant-in-Aid for Scientific Research (A) (No. 32678), COE Research (No. 13CE2002) and an aid fund from ENERGIA.

\end{document}